\documentclass[prd,twocolumn,aps]{revtex4}
\usepackage{amsmath}
\usepackage{graphicx}

\def\bea{\begin{eqnarray}}
\def\eea{\end{eqnarray}}
\def\beqa{\begin{equation}}

\def\eeqa{\end{equation}}

\def\be{\begin{equation}}
\def\ee{\end{equation}}

\def\5{\overline 5}

\newcommand{\prt}{\partial}

\def\beq{\begin{equation}}
\def\eeq{\end{equation}}

 \newcommand{\lsim}{\mbox{\raisebox{-1.ex}{$\stackrel
 {\textstyle<}{\textstyle \sim}$}}}
 \newcommand{\square}{\kern1pt\vbox{\hrule height
1.2pt\hbox{\vrule width 1.2pt\hskip 3pt
\vbox{\vskip 6pt}\hskip 3pt\vrule width 0.6pt}\hrule
height 0.6pt}\kern1pt}

\begin{document}
\title{Constraints on Dirac-Born-Infeld type dark energy models
\\from varying alpha}
\author{Mohammad R.~Garousi$^1$,
M.~Sami$^2$ and Shinji Tsujikawa$^3$}
\address{$^1$ Department of Physics, Ferdowsi University, P.O.Box 1436,
Mashhad, Iran\\ and\\
Institute for Studies in Theoretical Physics and Mathematics IPM\\
P.O.Box 19395-5531, Tehran, Iran}
\address{$^2$ IUCAA, Post Bag 4, Ganeshkhind,
Pune 411 007, India}
\address{$^3$ Department of Physics, Gunma National College of
Technology, 580 Toriba, Maebashi, Gunma 371-8530,
Japan}

\date{\today}

\begin{abstract}

We study the variation of the effective fine structure constant alpha
for Dirac-Born-Infeld (DBI) type dark energy models.
The DBI action based on string theory naturally gives
rise to a coupling between gauge fields and a scalar field
responsible for accelerated expansion of the universe.
This leads to the change of alpha due to a dynamical evolution of
the scalar field, which can be compatible with
the recently observed cosmological data around the redshift
$\tilde{z} \lesssim 3$. We place constraints on several different
DBI models including exponential, inverse
power-law and rolling massive scalar potentials.
We find that these models can satisfy the varying alpha constraint
provided that mass scales of the potentials are fine-tuned.
When we adopt the mass scales which are motivated by string theory,
both exponential and inverse power-law potentials give unacceptably
large change of alpha, thus ruled out from observations.
On the other hand the rolling massive scalar potential
is compatible with the observationally allowed variation of alpha.
Therefore the information of varying alpha provides a powerful way
to distinguish between a number of string-inspired
DBI dark energy models.

\end{abstract}
\maketitle
\vskip 1pc
\maketitle \vskip 1pc

\section{Introduction}

One of the most remarkable discoveries in cosmology is
attributed to the late time acceleration of universe which is supported
by supernova observations \cite{rp} and receives
independent confirmation from CMB and
galaxy clustering studies.
The current acceleration may be accounted for by supplementing
an exotic form of energy density with a negative pressure,
popularly known as dark energy (see Refs.~\cite{review} for reviews).
It was earlier thought that this could originate from a cosmological
constant, but the idea is fraught with an extreme fine-tuning problem.

This problem is alleviated in scalar-field models in which
the energy density of dark energy dynamically changes such that
it remains sub-dominant during the radiation and matter dominant eras
and becomes dominant at present.
In recent years, a wide variety of dark energy models
have been proposed, including Quintessence \cite{Paul},
K-essence \cite{Kes}, Chaplygin gas \cite{gas},
modifications of gravity \cite{modi}, Born-Infeld scalars
(rolling tachyon \cite{tach}, massive scalars \cite{GST}),
with the last one being originally motivated
by string theoretic ideas \cite{TA}.
The common feature of these models is that they operate
through an undetermined field potential
which in principal can incorporate any priori assigned cosmological
evolution, thus lacking predictive power at the fundamental
level \cite{tach}. These models should be judged by their
physical implication and by the generic features which arise in them.
There is tremendous degeneracy in this description and it is therefore
important to find other physical criteria which can constrain these models.
One such criterion can be provided by the variation of the effective fine
structure constant alpha on cosmological scales.

The old idea of time-varying fundamental physical constants \cite{dirac}
has recently attracted much attention in cosmology
(see Ref.~\cite{Uzan} for review).
In fact the Oklo natural fission reactor \cite{oklo} found the variation
of alpha
with the level $-0.9\times 10^{-7}< \Delta \alpha/\alpha<1.2\times 10^{-7}$
at the redshift $\tilde{z} \sim 0.16$.
The absorption line spectra of distance quasars \cite{savedov,wolf,murphy}
suggests
that  $\Delta \alpha/\alpha =(-0.574\pm 0.102) \times 10^{-5}$
for $0.2<\tilde{z}<3.7$ \cite{dzuba}.
The recent detailed analysis of high quality
quasar spectra \cite{chand} gives the lower variation of alpha,
$\Delta \alpha/\alpha=(-0.06\pm0.06)\times 10^{-5}$ over the
redshift range $0.4<\tilde{z}<2.3$.

It is well known that the interaction of scalar fields with gauge fields
can lead to the variation of the effective fine structure constant.
Typically the coupling is chosen to be arbitrary and ad hoc in
most of the scalar-field dark energy
models \cite{Bek,LS,DZ,Chiba,LV,SBM,OP,Wett,AG,CNP,NJ04,PBB,LLN,BK04,FM,LOP},
reflecting the fact that the coupling of a quintessence field $\phi$ to
matter
and radiation is not fixed by the standard model of particle physics
(see Refs.~\cite{alphaindustry} on the proposal of the least coupling
principal).
Assuming specific forms of the interaction and tuning the coupling parameters
along with the appropriate choice of field potentials may lead to a
desired change
in the effective fine structure constant.
However it looks unsatisfactory to work with several arbitrary functions and
parameters to produce a physically well-motivated result;
such models have a limited predictive power.
It is, therefore, necessary to look for the possibilities of obtaining  these
couplings from a fundamental theory which could fix
the above mentioned arbitrariness in the models.
Such couplings can be consistently derived from an effective D-brane action.

D-branes are extended dynamical objects in string theory on which
the end points of open  strings live. Their tree level action is
given by the Dirac-Born-Infeld (DBI) type action which contains
gauge fields and  scalar fields (tachyons \cite{TA}, massless
scalars \cite{tseytlin} and massive scalars \cite{mg}) . The DBI
action naturally gives rise to the coupling of the Born-Infeld
scalars with gauge fields, which can account for the
variation of the electromagnetic coupling over cosmological time
scales. As extensively studied in
Refs.~\cite{tach,AF03,AL04,CGST} the tachyon field might be
responsible for accelerated expansion of the universe at late
times (see Refs.~\cite{tachpapers} for the cosmological general
dynamics of tachyon). In this paper we study the variation of
alpha in the presence of the Born-Infeld scalar coupled to gauge
fields and place constraints on the model parameters. In fact
this provides us a powerful way to distinguish several different
tachyon potentials.

The paper is organized as follows. In
section II we examine the DBI action to derive the form of the coupling
between the tachyon and gauge fields
and set up the frame work for our analysis.
Then we study the variation of alpha together with the background
cosmological
evolution for several different tachyon potentials--(i) exponential
(Sec.~III),
(ii) inverse power-law (Sec.~IV) and (iii) rolling massive scalar (Sec.~V).
We investigate the evolution of alpha both analytically and numerically for
arbitrary mass scales of the potentials and then proceed to the case in which
the mass scale is constrained by string theory.
We show that the information of varying alpha provides important constraints
on tachyon potentials.
Section VI concludes our results.

\section{DBI Model}

We start with a Dirac-Born-Infeld type effective
4-dimensional action
\begin{equation}
\label{action0}
{\cal S}= -\int d^4x\,\tilde{V}(\varphi)
\sqrt{-\det(g_{\mu\nu}+
\partial_{\mu}\varphi\partial_{\nu}
\varphi+2\pi\alpha'F_{\mu \nu})}\,,
\end{equation}
where $\tilde{V}(\varphi)$ is the potential of a scalar field
$\varphi$ and  $F_{\mu \nu} \equiv 2\nabla_{[\mu}A_{\nu]}$ is a
Maxwell tensor with $A_\mu$ the four-potential. $\alpha'$ is
related to string mass scale $M_s$ via $\alpha'=M_s^{-2}$. We
are interested in a situation in which brane is located in a
ten-dimensional spacetime with a warped metric \cite{GST}
\begin{eqnarray}
\label{wmetric}
ds_{10}^2=\beta g_{\mu \nu}(x)dx^{\mu} dx^{\nu}+
\beta^{-1}\tilde{g}_{mn}(y)dy^mdy^n\,,
\end{eqnarray}
where $\beta$ is a warped factor. Note that the first term on the
r.h.s. of Eq.~(\ref{wmetric}) corresponds to the metric on the
brane.

The action (\ref{action0}) for this metric yields
\begin{eqnarray}
\label{action1}
{\cal S} &=& -\int d^4x\,\beta^2 \tilde{V}(\varphi) \nonumber \\
& & \times
\sqrt{-\det(g_{\mu\nu}+\beta^{-1}
\partial_{\mu}\varphi\partial_{\nu}\varphi+2\pi\alpha'
\beta^{-1} F_{\mu \nu})}.
\end{eqnarray}
Introducing new variables
\begin{eqnarray}
\label{trans}
\phi=\varphi/\sqrt{\beta}\,,~~~
V(\phi)=\beta^2 \tilde{V}(\sqrt{\beta}\phi)\,,
\end{eqnarray}
the action (\ref{action1}) can be written as
\begin{equation}
\label{action}
{\cal S}= -\int d^4x\,V(\phi)
\sqrt{-\det(g_{\mu\nu}+
\partial_{\mu}\phi\partial_{\nu}
\phi+2\pi\alpha' \beta^{-1} F_{\mu \nu})}\,.
\end{equation}
The warped metric (\ref{wmetric}) makes the mass scale on the brane
from  the string mass scale $M_s=1/\sqrt{\alpha'}$
to an effective mass which is $m_{\rm eff}=\sqrt{\beta}M_s$.
In what follows we shall consider cosmological dynamics and
the variation of alpha for the action (\ref{action}).

We adopt a spatially flat Friedmann-Robertson-Walker (FRW) metric
on the  brane with a scale factor $a(t)$ ($t$ is cosmic time). We
also account for the contributions of non-relativistic matter and
radiation, whose energy densities are $\rho_m$ and $\rho_r$,
respectively. Then the background equations of motion are
\begin{eqnarray}
\label{Hubble} & & \dot{H} =
-\frac{1}{2}\left( \frac{\dot{\phi}^2V}{\sqrt{1-\dot{\phi}^2}}+
\rho_m+\frac{4}{3}\rho_r\right)\,,\\
& & \frac{\ddot{\phi}}{1-\dot{\phi}^2}+3H\dot{\phi} +
\frac{V_\phi}{V}=0\,,\\
& &\dot{\rho}_m+3H\rho_m=0\,,\\
& & \dot{\rho}_r+4H\rho_r=0\,,
\label{phi}
\end{eqnarray}
together with a constraint equation for the Hubble rate
$H \equiv \dot{a}/a$ :
\bea
3M_p^2H^2 =
\frac{V(\phi)}{\sqrt{1-\dot{\phi}^2}}+\rho_m+\rho_r\,,
\eea
where $M_p$ is the reduced Planck mass ($M_p^{-2}=8\pi G$).
In the above equations we assumed the condition,
$|F_{\mu \nu}| \ll m^2_{\rm eff}$.

If the energy density of the field $\phi$ dominates at
late times and it leads to the acceleration of the universe,
we can employ the following slow-roll approximation:
\bea
\label{sapp}
3M_p^2H^2 \simeq V(\phi)\,,
~~~3H\dot{\phi} \simeq -\frac{V_\phi}{V}\,.
\eea
The slow-roll parameter for the DBI-type scalar field
is defined by \cite{tachpapers}
\bea
\label{spadef}
\epsilon \equiv -\frac{\dot{H}}{H^2}={M_p^2 \over 2}
\left(\frac {V_{\phi}}{V}\right)^2{ 1 \over V}\,.
\eea

We rewrite the above equations in an autonomous form.
Following Ref.~\cite{CLW}
we define the following dimensionless quantities:
\begin{eqnarray}
\label{Dquantity}
x \equiv \dot{\phi}=H\phi'\,,~~~
y \equiv \frac{\sqrt{V(\phi)}}{HM_p}\,,~~~
z \equiv \frac{\sqrt{\rho_r}}{\sqrt{3}HM_p}\,,
\end{eqnarray}
where a prime denotes the derivative with respect to
the number of $e$-folds, $N={\rm ln}\,a$.
Then we obtain the following equations
(see Refs.~\cite{AL04,CGST} for related works):
\begin{eqnarray}
\label{dotx}
& & x'=-(1-x^2)(3x-\lambda y)\,, \\
& & y'=\frac{y}{2}\left(-\lambda xy-\sqrt{1-x^2}\,y^2
+z^2+3\right) \,,\\
& & z'=z \left(-2-\frac12 \sqrt{1-x^2}\,y^2
+\frac12 z^2+\frac32\right)\,,\\
& &\lambda'=-\lambda^2 xy \left(\Gamma-\frac32
\right)\,,
\label{auto}
\end{eqnarray}
where
\begin{eqnarray}
\label{lam}
\lambda=-\frac{M_pV_\phi}{V^{3/2}}\,,~~~
\Gamma=\frac{VV_{\phi\phi}}{V_\phi^2}\,.
\end{eqnarray}
We note that $\lambda$ is related to $\epsilon$ by the relation
$\lambda^2=2\epsilon$. Therefore one has $|\lambda| \ll 1$ when the
slow-roll condition $\epsilon \ll 1$ is satisfied. We also define
\begin{eqnarray}
\Omega_\phi  &\equiv&
\frac{\rho_\phi}{\rho_{\rm cr}}
=\frac{y^2}{3\sqrt{1-x^2}}\,, \\
\Omega_r &\equiv& \frac{\rho_r}{\rho_{\rm cr}}=z^2\,,\\
\Omega_m &\equiv& \frac{\rho_m}{\rho_{\rm cr}}
=1-\frac{y^2}{3\sqrt{1-x^2}}-z^2\,,
\end{eqnarray}
where $\rho_{\rm cr} \equiv 3M_p^2H^2$ is a critical
energy density.
Note that these satisfy the constraint equation
$\Omega_\phi+\Omega_r+\Omega_m=1$.

The expansion of the action (\ref{action}) to second order of
the gauge field, for arbitrary metric, becomes
\begin{eqnarray}
{\cal S}& \simeq & \int d^4x\Big[-V(\phi)\sqrt{-\det(g_{\mu \nu} +\prt_\mu
\phi\prt_\nu \phi)} \nonumber \\
& & +\frac{(2\pi\alpha')^2V(\phi)}{4\beta^2}\sqrt{-g}
\,{\rm Tr}(g^{-1}Fg^{-1}F)\Big]\,. \eea
There are many other terms that
are the second order of the gauge field. They involve the derivative of
the field $\phi$ which we are not interested in.
Comparing the above action with the standard Yang-Mills action,
one finds that the effective fine-structure `constant' $\alpha$ is
\begin{eqnarray}
\alpha \equiv g_{\rm YM}^2=\frac{\beta^2M_s^4}{2\pi V(\phi)}\,.
\end{eqnarray}
The present value of fine structure constant is $\alpha=1/137$.
Since the potential energy of $\phi$ at present is estimated as
$V(\phi_0) \simeq 3H_0^2M_p^2$, one finds
\bea
\label{betas}
\beta^2 \simeq \frac{6\pi}{137}\left(\frac{H_0}{M_s}\right)^2
\left(\frac{M_p}{M_s}\right)^2\,.
\eea
If the string mass scale is the same order as the Planck mass
($M_s \simeq M_p$), we obtain
\bea
\label{warp}
\beta \simeq 10^{-61}\,,
\eea
where we used $H_0 \sim 10^{-42}$ GeV.
This property holds independent of the form of the
scalar-field potential.

The variation of $\alpha$ compared to the present value
$\alpha_0$ is given as
\begin{eqnarray}
\label{delal}
\frac{\Delta \alpha}{\alpha} \equiv \frac{\alpha-\alpha_0}
{\alpha_0}=\frac{V(\phi_0)}{V(\phi)}-1\,,
\end{eqnarray}
where $\phi_0$ is the present value of the field.
In the case of the exponential potential $V(\phi)=V_0e^{-\mu\phi}$,
we obtain
\begin{eqnarray}
\label{expdelal}
\frac{\Delta \alpha}{\alpha}
=e^{\mu (\phi-\phi_0)}-1
\simeq \mu (\phi-\phi_0)\,,
\end{eqnarray}
where the last equality is valid for $|\mu (\phi-\phi_0)| \ll 1$.
This corresponds to the choice (2.14) in Ref.~\cite{CNP}.
The authors in Ref.~\cite{CNP} chose the coupling of the
type $\Delta \alpha/\alpha \propto \Delta\phi$ phenomenologically
for {\it any} potentials of the field $\phi$.
In our case, however, this is dependent on the form of $V(\phi)$.

For the inverse power-law potential $V(\phi)=M^{4-n}\phi^{-n}$
considered in Refs.~\cite{tach,AF03,AL04,CGST}, one gets
\begin{eqnarray}
\label{dalinverse}
\frac{\Delta \alpha}{\alpha}=\left(\frac{\phi}{\phi_0}
\right)^n-1\,,
\end{eqnarray}
and for the massive rolling  scalar potential
$V(\phi)=V_0e^{\frac{1}{2}M^2\phi^2}$ considered in
Refs.~\cite{GST}, we have
\begin{eqnarray}
\frac{\Delta \alpha}{\alpha} =
e^{-\frac12 M^2 (\phi^2-\phi_0^2)}-1\,.
\label{varialp}
\end{eqnarray}

In subsequent sections we shall study the variation of alpha
for several different tachyon potentials and place constraints
on the parameters of the models.
Recently tachyon potentials are classified
in three ways \cite{CGST}--
(i) $\lambda={\rm const}$, (ii) $\lambda \to 0$ asymptotically
and (iii) $|\lambda| \to \infty$ asymptotically.
The case (i) corresponds to an inverse square potential
$V(\phi) \propto \phi^{-2}$, which gives the border of
acceleration and deceleration (see Refs.~\cite{tach,AF03,PT}).
The case (ii) gives rise to an accelerated expansion at late times.
An example is provided by inverse power-law potential
$V(\phi) \propto \phi^{-n}$ with $n<2$.
The potential $V(\phi)=V_0e^{\frac12 M^2\phi^2}$ of
a rolling massive scalar field also belongs to this class.
The case (iii) corresponds to a deceleration at late times,
but with a possible stage of transient acceleration.
The exponential potential $V(\phi)=V_0 e^{-\mu \phi}$
belongs to this class.

In Sec.~III we shall consider the exponential potential
as an example of the case (iii). In Sec.~IV the inverse
power-law potential $V(\phi) \propto \phi^{-n}$
with $0 \le n \le 2$ is studied as examples of
the cases (i) and (ii). The rolling massive scalar
potential $V(\phi)=V_0e^{\frac12 M^2\phi^2}$ is
discussed in Sec.~V, since this has somewhat different
property compared to other potentials in which
the field rolls down toward infinity.

\section{Exponential potentials}

We first consider the exponential potential
\begin{eqnarray}
V(\phi)=V_0e^{-\mu \phi}\,,
\label{exp}
\end{eqnarray}
where $\mu$ has a dimension of mass.  The exponential potential
may appear in the context of the D$_3$ anti-D$_3$ cosmology
\cite{KK1}. The tachyon potential for the coincident D$_3$
anti-D$_3$ is twice the potential  for  non-BPS D$_3$-brane
\cite{senmg}. The latter  is given by
$V(\phi)=2\beta^2T_3/\cosh(\sqrt{\beta}M_s\phi)$ \cite{Liu}, where
$\beta$ is a warp factor at the position of the D$_3$ anti-D$_3$
in the internal compact space, $T_3$ is the tension of branes.
Then the potential behaves as $V(\phi)\sim \beta^2 T_3
e^{-\sqrt{\beta}M_s \phi}$ at sufficiently late times 
($\phi \to \infty$), which has a correspondence
\begin{eqnarray}
\label{muexp}
V_0 = \beta^2 T_3\,,~~~
\mu = \sqrt{\beta}M_s\,,
\end{eqnarray}
in Eq.~(\ref{exp}).
Therefore $V_0$ and $\mu$ are not free parameters if we restrict
to string theory.
In what follows we will consider cosmological constraints on
the exponential potential (\ref{exp}) in general and then
proceed to the case in which the potential is motivated by
string theory.

By Eq.~(\ref{spadef}) the slow-roll parameter is given by
$\epsilon=\mu^2 M_p^2e^{\mu \phi}/2V_0$.
The accelerated expansion occurs for $\epsilon<1$,
but this stage eventually ends for
$\phi>\phi_f \equiv (1/\mu) {\rm log}\,(2V_0/\mu^2M_p^2)$.
Therefore the current acceleration of the universe corresponds to
transient quintessence for exponential potentials.
Let us employ the slow-roll approximation (\ref{sapp}) around
the redshift $\tilde{z} \lesssim {\cal O}(1)$ under the condition
that the energy density of the universe is dominated by $\rho_\phi$.
Then the slow-roll parameter $\epsilon$ is simply written as
\begin{eqnarray}
\label{epexpo}
\epsilon \simeq \frac{\mu^2}{6H_0^2}\,.
\end{eqnarray}
This means that accelerated expansion occurs
for $\mu \lesssim H_0$.

The evolution of the field $\phi$ is given by \cite{sami}
\begin{eqnarray}
\label{phiexp}
e^{-\mu \phi_0/2}-e^{-\mu \phi/2}=
\frac{\mu^2 M_p}{2\sqrt{3V_0}}(t-t_0)\,,
\end{eqnarray}
where $\phi_0$ and $t_0$ are the present values.
Under the condition with $|\mu (\phi-\phi_0)| \ll 1$,
which is actually required
from the observational constraint $|\Delta \alpha/\alpha| \ll 1$ in
Eq.~(\ref{expdelal}),
we have that $\phi-\phi_0 \simeq \mu M_p e^{\mu \phi/2}/\sqrt{3V_0}(t-t_0)$.
Since the redshift is given by $\tilde{z}\simeq -H_0(t-t_0)$ for small
$\tilde{z}$, we find that the time-variation of alpha is approximately
written as
\begin{eqnarray}
\label{delalexp}
\frac{\Delta \alpha}{\alpha} \simeq
-\frac{\mu^2}{3H_0^2}\tilde{z}\,.
\end{eqnarray}
In order to obtain $|\Delta \alpha/\alpha|=10^{-6}$-$10^{-5}$
around $\tilde{z}={\cal O}(1)$, the mass scale $\mu$ is
constrained to be $\mu/H_0=10^{-3}$-$10^{-2}$.
In this case the accelerated expansion actually occurs, since
$\epsilon$ is much smaller than 1 by Eq.~(\ref{epexpo}).

\begin{figure}
\begin{center}
\includegraphics[height=3.5in,width=3.5in]{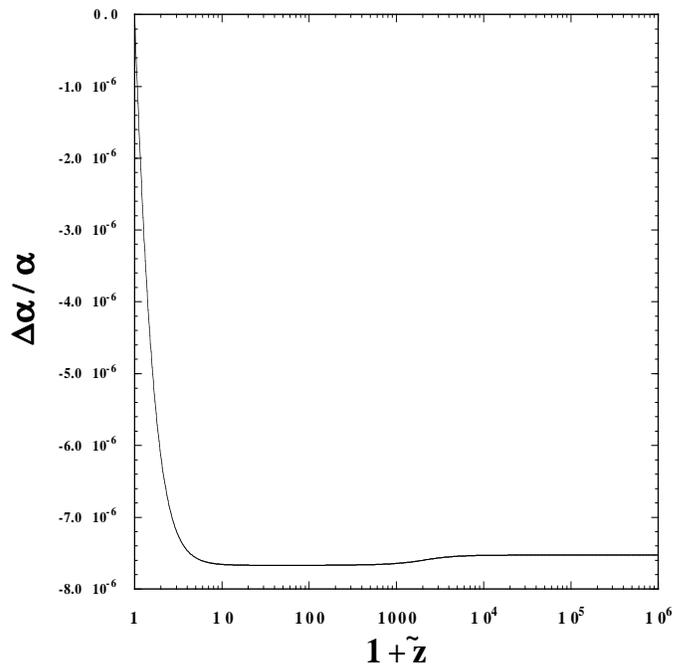}
\caption{
The variation of $\Delta \alpha/\alpha$
as a function of the redshift $\tilde{z}$
for the exponential potential (\ref{exp})
with $\mu=8.65 \times 10^{-3}H_0$.
Initial conditions are chosen to be
$x_i-1=-1.0 \times 10^{-12}$,
$y_i=2.5 \times 10^{-10}$, $z_i=0.99$ and
$\lambda=5.0 \times 10^{-3}$ at $\tilde{z}=10^6$.
The present epoch ($\tilde{z}=0$) corresponds to
$\Omega_\phi \simeq 0.7$ and $\Omega_m \simeq 0.3$.
}
\label{delalphaexp}
\end{center}
\end{figure}

\begin{figure}
\begin{center}
\includegraphics[height=3.5in,width=3.5in]{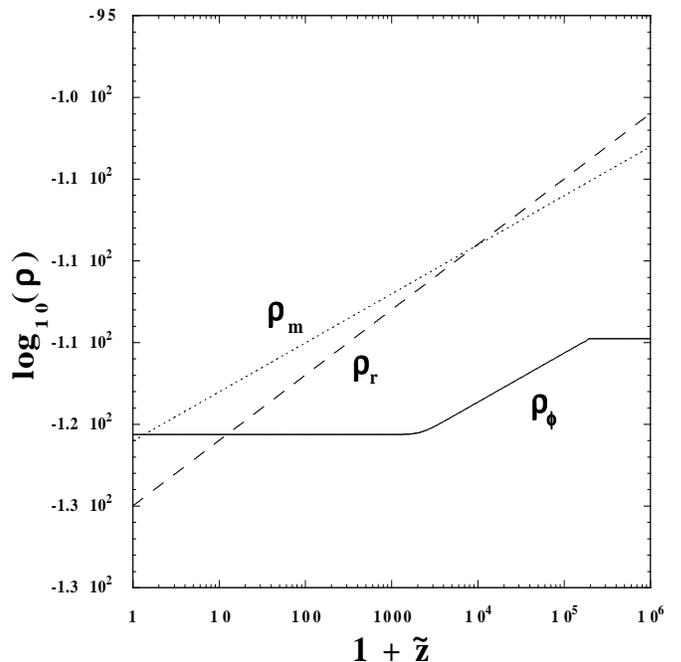}
\caption{The evolution of the energy densities
$\rho_\phi$, $\rho_r$ and $\rho_m$
for the exponential potential (\ref{exp})
with $\mu=8.65 \times 10^{-3}H_0$.
Initial conditions are the same as in Fig.~\ref{delalphaexp}.
}
\label{logrho}
\end{center}
\end{figure}

We need to caution that slow-roll analysis we used is not
necessarily valid in presence of the background energy density. In order
to confirm the validity of the above analytic estimation, we
numerically solved the background equations
(\ref{dotx})-(\ref{auto}) together with Eq.~(\ref{delal}). Figure
\ref{delalphaexp} shows one example of the variation of alpha for
$\mu=8.65 \times 10^{-3}H_0$. Initial conditions of the variables
$x$, $y$ and $z$ are chosen so that a viable cosmological
evolution can be obtained starting from the radiation dominant
era. We begin integrating around from the redshift
$\tilde{z}=10^6$ with $\dot{\phi}^2$ very close to 1. In
Fig.~\ref{logrho} we find that $\rho_\phi$ decreases similarly to
$\rho_m$ around $10^3 \lesssim \tilde{z} \lesssim 10^5$, which
comes from the fact that the tachyon behaves as a pressureless
dust for $\dot{\phi}^2 \sim 1$. This is followed by a stage with
slowly changing energy density $\rho_\phi$ that behaves as an
effective cosmological constant. The tachyon evolves very slowly
during this stage ($\dot{\phi}^2 \ll 1$), which leads to the
accelerated expansion once the energy density $\rho_\phi$ becomes
dominant. In the case of Fig.~\ref{delalphaexp} we have
$\Omega_\phi=0.7$ and $\Omega_m=0.3$ at present ($\tilde{z}=0$)
with $\epsilon=\lambda^2/2 \ll 1$. The universe will eventually
enter a phase with a decelerated expansion in future after
$\lambda$ grows of order unity.

In Fig.~\ref{delalphaexp} we find
$|\Delta \alpha/\alpha| \simeq 10^{-6}$-$10^{-5}$ around the redshift
$\tilde{z}={\cal O}(1)$, thus showing the validity of
our analytic estimation based on the slow-roll approximation.
Since one has $\Delta \alpha/\alpha \simeq -2\epsilon \tilde{z}$
by Eqs.~(\ref{epexpo}) and (\ref{delalexp}),
the slow-roll condition, $\epsilon \ll 1$,
is crucially important to provide an appropriate level of the variation
of the effective fine structure constant ($|\Delta \alpha/\alpha| \ll 1$).
It is intriguing that the condition for an accelerated expansion is
compatible with that for varying alpha around 
$\tilde{z} \lesssim {\cal O}(1)$.
We note that pure cosmological constant does not give
rise to any variation of alpha. In this sense the information of
varying alpha is important to distinguish between the cosmological constant
and the DBI dark energy model.

If we consider the exponential potential motivated by
string theory, the mass $\mu$ may be replaced by
$\mu=\sqrt{\beta}M_s$ [see Eq.~(\ref{muexp})].
Then by Eqs.~(\ref{betas}) and (\ref{delalexp})
we find
\begin{eqnarray}
\frac{\Delta \alpha}{\alpha} \simeq
-\sqrt{\frac{2\pi}{411}} \frac{M_p}{H_0} \tilde{z}
\simeq -10^{59}\,\tilde{z}\,.
\end{eqnarray}
This gives $|\Delta \alpha/\alpha| \gg 1$ for
$\tilde{z}={\cal O}(1)$, which completely contradicts
with observational values. Therefore the mass scale $\mu$
given in Eq.~(\ref{muexp}) is too large to account for the
small variation of alpha ($\Delta \alpha/\alpha| \ll 1$).
This means that exponential potentials of the tachyon
are ruled out from the information of varying alpha
if we adopt the mass scale $\mu$ motivated by string
theory.

\section{Inverse power-law potentials}

The model based upon exponential potentials involves
very small mass scales ($\mu \ll H_0$) and suffers
from a severe fine tuning problem. We now consider a class
of inverse power-law potentials for which the problem of
the fine tuning may be considerably reduced.
The potentials are of the form:
\begin{eqnarray}
V(\phi)=M^{4-n}\phi^{-n}\,,
\label{powerlaw}
\end{eqnarray}
where $M$ has a dimension of mass.

Here again we should emphasize that $M$  is not  necessarily
a free parameter if we restrict to string theory.
If ones takes the tachyon potential
$\tilde{V}(\varphi)=M_s^{4-n} \varphi^{-n}$ in the original
DBI action (\ref{action0}), we obtain the potential
(\ref{powerlaw}) in the action (\ref{action}) with
\begin{eqnarray}
M=\sqrt{\beta}M_s\,.
\label{powerlawmass}
\end{eqnarray}
We shall first consider arbitrary values of $M$ and then
proceed to the case in which the relation (\ref{powerlawmass})
is used.

As shown in Ref.~\cite{AF03,CGST}
the accelerated expansion occurs at a later stage
for $0 \le n \le 2$.
Employing the slow-roll approximation (\ref{sapp}),
we get the following solution
\begin{eqnarray}
A_n(t-t_0)&=& \nonumber \phi^{(4-n)/2}-\phi_0^{(4-n)/2} \,, \\
A_n&=&\frac{n(4-n)M^{(n-4)/2}M_p}{2\sqrt{3}}\,,
\label{An}
\end{eqnarray}
where $t_0$ and $\phi_0$ are present values.
Around the region $t \sim t_0$ we can expand the solution as
\begin{eqnarray}
\phi=\phi_0 \left[1+B_n(t-t_0)\right]\,,~~
B_n=\frac{nM_p}{\sqrt{3}(\phi_0 M)^{(4-n)/2}}\,.
\end{eqnarray}
In order for this expansion to be valid, we require
$|B_n(t-t_0)| \ll 1$.

Then Eq.~(\ref{dalinverse}) gives
\begin{eqnarray}
\frac{\Delta \alpha}{\alpha}=nB_n(t-t_0)
=-\frac{n^2M_p}{\sqrt{3}(\phi_0 M)^{(4-n)/2}H_0}\tilde{z}\,.
\label{Deal}
\end{eqnarray}
Using the slow-roll approximation
$3H_0^2M_p^2 \simeq M^{4-n}\phi_0^{-n}$ around
$\tilde{z} \lesssim {\cal O}(1)$, we find
\begin{eqnarray}
\label{delalcon}
\frac{\Delta \alpha}{\alpha}=
-\frac{n^2}{3^{1-2/n}}\left(\frac{M}{M_p}
\right)^{2-8/n} \left(\frac{H_0}{M_p}\right)^{4/n-2}
\tilde{z}\,.
\end{eqnarray}
For example one has $\Delta \alpha/\alpha=
-4(M/M_p)^{-2}\tilde{z}$ for $n=2$.
By Eq.~(\ref{delalcon}) we can constrain the mass scale
$M$ by using the information of varying alpha:
\begin{eqnarray}
\frac{M}{M_p}=\left[\biggl|\frac{\Delta \alpha}
{\alpha}\biggr| \frac{1}{n^2\tilde{z}}
\left(\frac{M_p}{\sqrt{3}H_0}\right)^
{(4-2n)/n}\right]^{n/(2n-8)}\,.
\label{mscale}
\end{eqnarray}

Let us first consider the case of $n=2$ whose cosmological
evolution was investigated in Ref.~\cite{tach,AL04}.
When we use the constraint $|\Delta \alpha/\alpha|~\lsim~10^{-6}$
for $\tilde{z}={\cal O}(1)$, we have $M~\gtrsim~10^{3}M_p$
by Eq.~(\ref{mscale}).
Such a large mass is obviously problematic since we expect
general relativity itself to break down in such a regime.
In order to obtain the mass scale $M$ that is smaller than
$M_p$, $\Delta \alpha/\alpha$ needs to be much greater than unity,
thus incompatible with observations.

The problem of the super-Planckian mass scale can be
circumvented by considering the class of less steeper potentials
with $n$ smaller than 2.
In Fig.~\ref{mass} we plot $M/M_p$
as a function of $n$ for several different values of
$|\Delta \alpha/\alpha|$ at $\tilde{z}=1$.
The allowed mass scale corresponds to the region
which is above the borders shown in this figure.
We then have $M \lesssim M_p$ for $n \lesssim 1.9$
and $M/M_p \sim 10^{-19}$ for $n=1$.
The mass $M$ has a minimum around $n=0.57$.
In the limit $n \to 0$ one has $M \to M_p$
by Eq.~(\ref{mscale}).
This corresponds to a cosmological constant $M^4$
with a Planck energy density.
Therefore the constraint from varying alpha gives
high energy scales around $n=0$, which does not
provide a realistic dark energy scenario.

\begin{figure}
\begin{center}
\includegraphics[height=3.5in,width=3.5in]{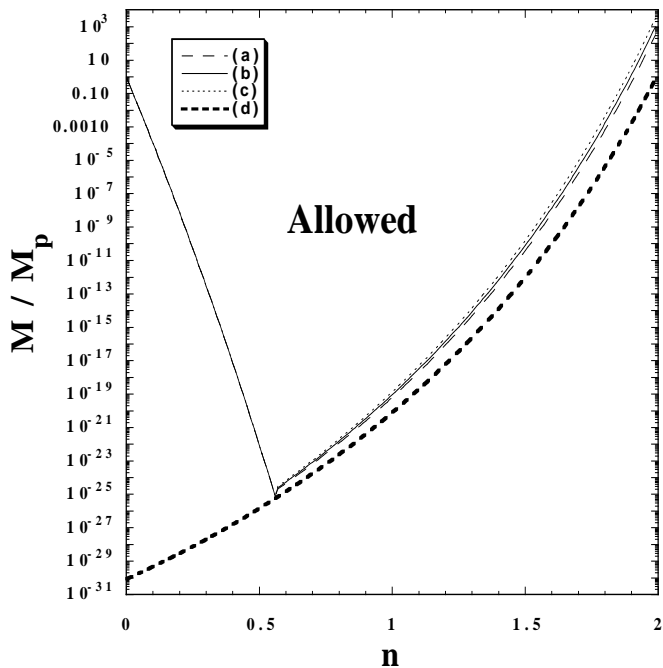}
\caption{
The mass scale $M/M_p$ determined by the varying alpha
constraint in terms of the function of the power $n$.
Each case corresponds to
(a) $|\Delta \alpha/\alpha|=10^{-5}$,
(b) $|\Delta \alpha/\alpha|=10^{-6}$, and
(c) $|\Delta \alpha/\alpha|=10^{-7}$ at $\tilde{z}=1$.
The curve (d) is the bound which is determined
by the condition for accelerated expansion,
see Eq.~(\ref{Mcons}).}
\label{mass}
\end{center}
\end{figure}

Let us consider the evolution of $\Delta \alpha/\alpha$
together with the background cosmological dynamics.
The slow-roll parameter for the field $\phi$ is
\begin{equation}
\label{slowpara}
\epsilon =\frac{n^2}{2} \left(\frac{M_p}{M}\right)^2
\frac{1}{(\phi M)^{2-n}}\,.
\end{equation}
This is constant for $n=2$, i.e., $\epsilon=2(M_p/M)^2$.
In order to get an accelerated expansion at late times,
we require $M>\sqrt{2}M_p$ in this case.
The mass scale which is constrained from
the information of varying alpha ($M/M_p \gtrsim 10^3$)
satisfies the condition for acceleration.

When $n<2$ the condition for accelerated expansion,
$\epsilon<1$, yields
\begin{equation}
\phi M>\left(\frac{n}{\sqrt{2}} \frac{M_p}{M}
\right)^{2/(2-n)}\,.
\label{phiM}
\end{equation}
The initial value of the field in the radiation dominant
epoch needs to be chosen so that it satisfies the condition
(\ref{phiM}) at late times.
For example one has $\phi M>5.0 \times 10^{37}$ for $n=1$
and $M/M_p=10^{-19}$.
We can place another constraint on the mass scale $M$.
The present potential energy is approximated by
$V(\phi_0)=M^4/(\phi_0 M)^n \simeq \rho_c \simeq
10^{-47} {\rm GeV}^4$ with $\phi_0$ satisfying
Eq.~(\ref{phiM}).
Then we obtain the relation
\begin{equation}
\frac{M}{M_p}>\left[\left(\frac{\rho_c}{M_p^4}\right)^{1-
\frac{n}{2}}\left(\frac{n}{\sqrt{2}}\right)^n\right]^{1/(4-n)}\,.
\label{Mcons}
\end{equation}

In Fig.~\ref{mass} this bound is plotted as a curve (d).
We find that the varying alpha bound (\ref{mscale}) provides
a severer constraint compared to (\ref{Mcons}) coming
from the condition of accelerated expansion.
Therefore it is important to take into account the information
of varying alpha when we constrain the inverse power-law potential.

We shall numerically study the evolution of $\Delta \alpha/\alpha$
in order to confirm the analytic estimation based on the slow-roll
approximation. First of all, we checked that the inverse square
potential ($n=2$) can explain the required variation of $\alpha$
($|\Delta \alpha/\alpha|=10^{-6}$-$10^{-5}$) provided that the
mass $M$ is much larger than the Planck mass.
However this mass scale is unacceptably large
from the viewpoint of the validity of general relativity.

Let us proceed to the case with $n<2$. By Eq.~(\ref{slowpara})
the slow-roll parameter decreases as the field evolves toward
larger values. We can consider two situations which lead to
the acceleration of the universe at late times.
The first is the case in which the slow-roll condition $\epsilon \ll 1$
is satisfied even for $\tilde{z}>{\cal O}(1)$ [we call this the case (a)].
In this case the field $\phi$ evolves slowly ($\dot{\phi}^2 \ll 1$)
during the transition from the
matter-dominant era to the scalar field dominant era.
The second corresponds to the case in which the slow-roll
parameter is larger than 1 for $\tilde{z} \gtrsim 1$
but becomes less than 1 for $\tilde{z} \lesssim 1$
[we call this the case (b)].
Since the transition from the non slow-roll phase to the slow-roll stage
occurs around $\tilde{z} \sim 1$ in this case, one can not necessarily employ
the approximation (\ref{sapp}) in this region.

\begin{figure}
\begin{center}
\includegraphics[height=3.5in,width=3.5in]{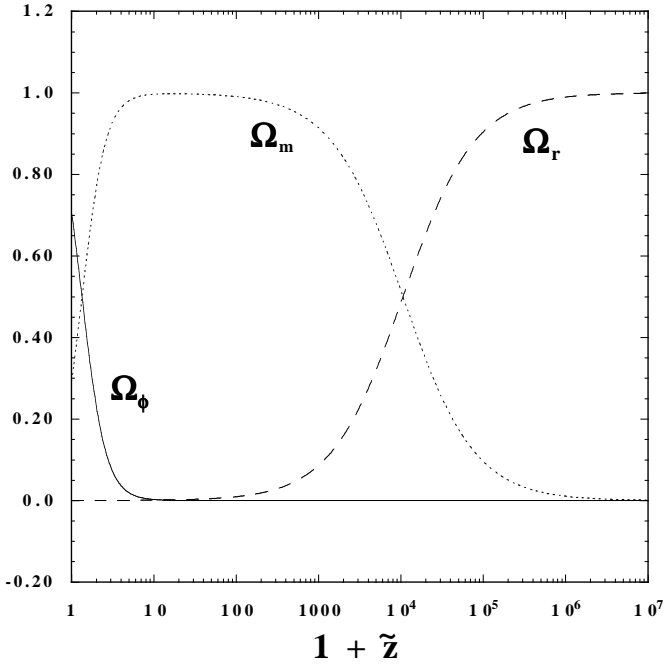}
\caption{
The evolution of $\Omega_\phi$, $\Omega_m$ and
$\Omega_r$ as a function of the redshift $\tilde{z}$
for the inverse power-law potential with
$n=1$ and $M=3.79 \times 10^{-19}$.
This case satisfies $\Omega_\phi=0.7$ and
$\phi_0M=1.50 \times 10^{42}$ at $\tilde{z}=0$.
This cosmological evolution corresponds to the case (a)
in which the slow-roll condition $\epsilon \ll 1$ is
satisfied for $\tilde{z}>{\cal O}(1)$.}
\label{Omega}
\end{center}
\end{figure}

\begin{figure}
\begin{center}
\includegraphics[height=3.5in,width=3.5in]{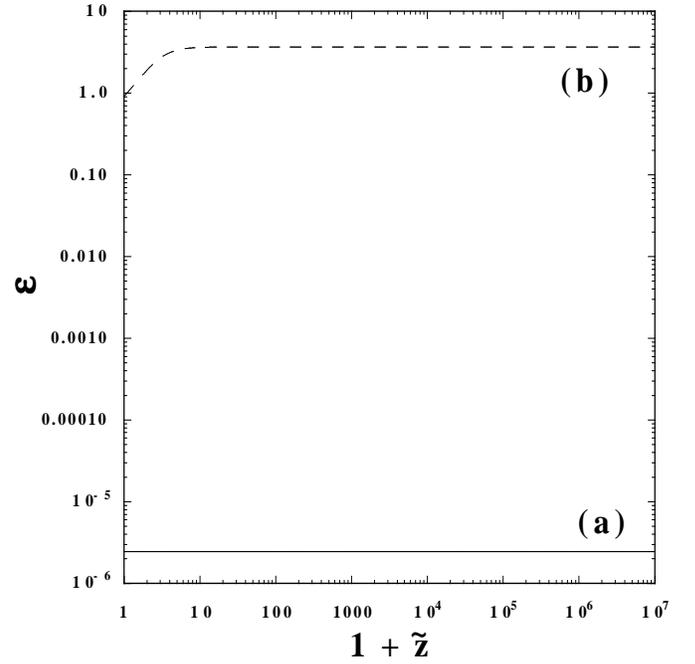}
\caption{
The evolution of the slow-roll parameter $\epsilon$
for the inverse power-law potential with
$n=1$ and $M=3.79 \times 10^{-19}$.
Each case corresponds to
(a) $\phi_0M=1.50 \times 10^{42}$ and
(b) $\phi_0M=3.98 \times 10^{36}$ at $\tilde{z}=0$.
In the case (a) $\epsilon$ is nearly constant with $\epsilon \ll 1$.
Meanwhile $\epsilon$ becomes smaller than 1 only near to the present
in the case (b).}
\label{epsilon}
\end{center}
\end{figure}

\begin{figure}
\begin{center}
\includegraphics[height=3.5in,width=3.5in]{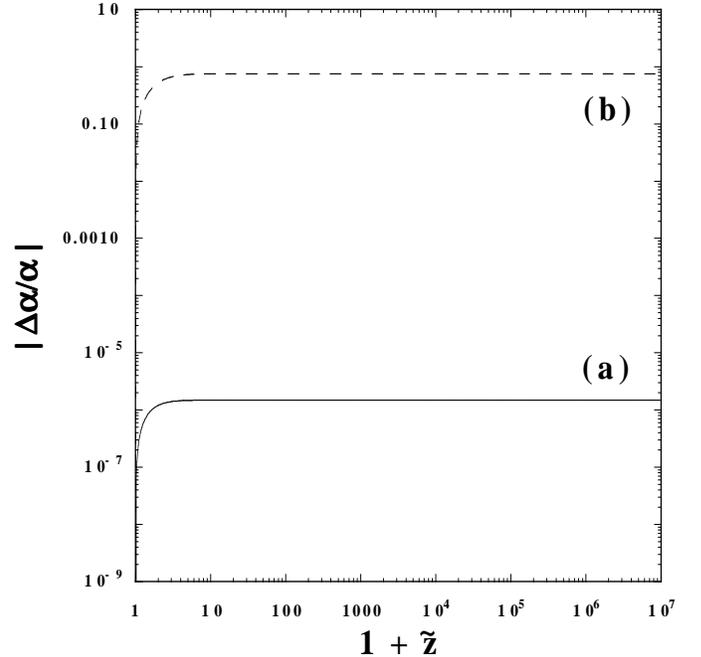}
\caption{
The variation of alpha as a function of the redshift
$\tilde{z}$ with same model parameters and initial conditions
as in Fig.~\ref{epsilon}.
The case (a) shows a small variation of alpha, whereas the
case (b) gives large values of $|\Delta \alpha/\alpha|$.
}
\label{Delal2}
\end{center}
\end{figure}

In both cases one can obtain a viable cosmological
evolution which reaches $\Omega_\phi \simeq 0.7$ and
$\Omega_m \simeq 0.3$ at present.
In Fig.~\ref{Omega} we plot the evolution of
$\Omega_\phi$, $\Omega_m$ and $\Omega_r$
for $n=1$ and $M=3.79 \times 10^{-19}$ with
the present value $\phi_0M=1.50 \times 10^{44}$
[corresponding to the case (a)].
In this case the slow-roll parameter
$\epsilon$ is much smaller than unity and is nearly constant
even for $\tilde{z}>{\cal O}(1)$, as plotted in Fig.~\ref{epsilon}.
When $\epsilon$ becomes less than unity
only near to the present [the case (b)],
we numerically checked that a viable cosmological solution that
leads to the late-time acceleration can be obtained as well.

We can distinguish the above two different cases
by having a look at the evolution of the effective fine structure constant.
Figure \ref{Delal2} shows that $|\Delta \alpha/\alpha|$ is kept to
be small ($|\Delta \alpha/\alpha| \sim 10^{-6}$) even for
$\tilde{z}<{\cal O}(1)$ in the case (a).
This reflects the fact that the field $\phi$ evolves very slowly
because of a small slow-roll parameter ($\epsilon \ll 1$).
On the other hand the situation is different in the case (b).
Since the evolution of the field $\phi$ is not
described by a slow-roll for $\tilde{z} \gtrsim 1$,
this leads to a larger variation of alpha
compared to the one derived by the slow-roll analysis.
This property is clearly seen in Fig.~\ref{Delal2}.

The above results indicate that the field $\phi$ needs to satisfy
the condition (\ref{phiM}) in the matter-dominant era
prior to the accelerating phase
in order for the variation of $\alpha$ to be compatible
with observations. Together with the minimum mass
scale $M$ determined by Eq.~(\ref{mscale}),
the information of varying alpha places a
constraint on the initial condition of $\phi$.
In the case (a) which gives
$|\Delta \alpha/\alpha|=10^{-6}$-$10^{-5}$ around
$\tilde{z}={\cal O}(1)$ the variation of alpha is small even
for $\tilde{z} \gg 1$, thus satisfying the nucleosynthesis
constraint $|\Delta \alpha/\alpha| <2 \times 10^{-8}$
around $\tilde{z}=10^8$-$10^{10}$ \cite{Ber} (see Fig.~\ref{Delal2}).

Let us finally consider the case in which the mass scale is
constrained by Eq.~(\ref{powerlaw}) from string theory.
Figure \ref{mass} indicates that $M/M_p>10^{-25}$
from the requirement of varying alpha.
Using Eqs.~(\ref{warp}) and (\ref{powerlaw}) we find that
the string mass scale is unacceptably large, i.e.,
$M_s/M_p \gtrsim 10^5$. In order to obtain $M_s \lesssim M_p$
we require the condition $|\Delta \alpha/\alpha| \gg 1$, which is
incompatible with observations.
Therefore the inverse-power law potential is disfavored
from the information of varying alpha when we use the
mass scale $M$ motivated by string theory.

\section{Rolling massive scalar potential}

Let us finally study the rolling massive scalar potential \cite{GST}
\begin{eqnarray}
\label{massivero}
V(\phi)=V_0e^{\frac12 M^2\phi^2}\,,
\end{eqnarray}
where $V_0$ and $M$ are constants.
If one considers the potential $V(\varphi)=T_3
e^{\frac12 M_s^2\varphi^2}$ in the original action (\ref{action0}),
one obtains the potential (\ref{massivero}) in the action (\ref{action})
with correspondence
\begin{eqnarray}
\label{V0M}
V_0 =\beta^2 T_3\,,~~~M =\sqrt{\beta}M_s\,,
\end{eqnarray}
The cosmological dynamics for this potential
was investigated in the context of inflation \cite{GST}
and dark energy \cite{CGST}. In both cases we require
small warp factor $\beta~(\ll 1)$ to satisfy
observational constraints.

In our scenario we recall that the warp factor is constrained
by Eq.~(\ref{betas}) from the present value of alpha.
In this case it is easy to confirm that the potential
(\ref{massivero}) with $V_0=\beta^2 T_3$ and $|\phi M| \ll 1$
can account for the the present value of the Hubble constant $H_0$,
provided that the string mass scale $M_s$ is the same order as
$T_3^{1/4}$. In what follows we shall study a situation in
which the field evolves close to the potential
minimum at present ($|\phi_0M| \ll 1$).

The evolution equation for the massive DBI scalar is
\begin{eqnarray}
\label{ddphieq}
\frac{\ddot{\phi}}{1-\dot{\phi}^2}+3H\dot{\phi}
+M^2\phi=0\,.
\end{eqnarray}
The Hubble rate at present is approximately given by
$H_0 \simeq \sqrt{V_0/(3M_p^2)}$. We wish to
consider the case with $V_0$ and $M$ given by
Eq.~(\ref{V0M}).
Then the mass $M$ is much larger than $H_0$
as long as $M_s$ is not too much smaller than $M_p$,
so the friction term is negligible around the potential minimum.
When $\dot{\phi}^2 \ll 1$ the solution for
Eq.~(\ref{ddphieq}) may be given by
\begin{eqnarray}
\label{Phieq}
\phi \simeq \Phi \cos(Mt)\,,
\end{eqnarray}
where $\Phi$ is the field amplitude.
In fact we numerically checked that the field $\phi$
oscillates as in Eq.~(\ref{Phieq}) for $M \gg H_0$
and $\dot{\phi}^2 \ll 1$.

In the oscillatory regime, the condition for acceleration for
the DBI scalar is
\begin{equation}
\langle \dot{\phi}^2 \rangle=M^2 \Phi^2 \frac{1}{T}
 \int_0^T{\sin^2(M t) {\rm d}t} < 2/3 \,,
\end{equation}
where $T$ is the period of oscillations.
This gives
\begin{eqnarray}
M^2 \Phi^2<4/3\,.
\end{eqnarray}
Let us consider the situation with $M^2 \Phi^2<4/3$.
Then by Eq.~(\ref{varialp}) the variation of $\alpha$ is
approximately given by
\begin{eqnarray}
\frac{\Delta\alpha}{\alpha}
\simeq -\frac12 M^2 (\phi^2-\phi_0^2)\,.
\end{eqnarray}
Since $|\phi^2-\phi_0^2| \lesssim \Phi^2$, we obtain
\begin{eqnarray}
\left|\frac{\Delta\alpha}{\alpha}\right| \lesssim
\frac12 M^2 \Phi^2 \,.
\end{eqnarray}

It is possible to have $|\Delta \alpha/\alpha| \sim 10^{-6}$-$10^{-5}$
if $|M\Phi|$ is of order $10^{-3}$-$10^{-2}$.
We recall that the condition for acceleration is satisfied in this case.
When $M \gg H_0$ the time scale of the oscillation of the field $\phi$
is much smaller than the cosmological time corresponding to the
redshift $\tilde{z}={\cal O}(1)$.
For the choice (\ref{V0M}), $M$ is much larger than the Hubble
rate $H$ around $\tilde{z}={\cal O}(1)$, which means that
the above estimation for alpha is valid even for
$\tilde{z}={\cal O}(1)$.
The field oscillates coherently for many times
while the universe evolves from $\tilde{z}={\cal O}(1)$
to $\tilde{z}=0$. In this case we have an interesting possibility
to explain the oscillation of alpha which is actually seen
in observational data \cite{chand,FM}.

In Ref.~\cite{GST} the potential (\ref{massivero}) was used for
the inflation in early universe, in which case the warp factor is
constrained to be $\beta \sim 10^{-9}$ from the COBE
normalization (in this scenario the problem of reheating is
overcome by accounting for a negative cosmological constant that
may come from the stabilization of the modulus). This warp factor
is very different from the one given in Eq.~(\ref{warp}). In the
so-called KKLT scenario \cite{KKLT}, one may try to link the
quintessential anti-D3 brane with the primordial inflationary
anti-D3 brane. To this end, one should think of a mechanism which
explains the reduction of $\beta$ from $\beta\sim 10^{-9}$ to
$\beta\sim 10^{-61}$ at some time after a reheating epoch. The
warp factor at the tip of the Klebanov-Strassler throat in which
the anti-D3 branes live is given by \cite{GKP}
\bea
\beta \sim \exp \left( -\frac{4\pi {\cal N}}{3g_s {\cal M}} \right)\,,
\eea
where $g_s$ is the string coupling constant, and the integers
${\cal M}$, ${\cal N}$ denote the R-R and NS-NS three form
fluxes, respectively, in the Calabi-Yau manifold of the compact
space. The warp factor has its minimum value at this point in the
KS throat. This minimum  can be extremely small for suitable
choice of fluxes. The RR flux  annihilation \cite{kach}  may  then
explain the reduction of the warp factor at the tip of the KS
throat.

We have found that the effective mass on anti-D3 branes is
$m_{\rm eff}=\sqrt{\beta}M_s\sim 10^{-12}\,{\rm GeV}$. This makes
all massive excitations of the branes to be quite light. However,
massive strings stretching between two nearly coincident such
light branes which may play the role of W-bosons need not to be
light. In fact the mass of stretched open string between two
branes with separation $\ell$ and in the warped metric
Eq.~(\ref{wmetric}) is given by
$M^2_W=\ell^n\ell^mg_{nm}=\beta^{-1}\ell^2\sim 10^{61}\ell^2$.
For small enough $\ell$ one can find a suitable W-boson mass.

\section{Conclusions}

In this paper we have studied the variation of the
electromagnetic coupling in the frame work of
Dirac-Born-Infeld (DBI) type dark energy models.
Since the Born-Infeld scalar field is generally coupled to gauge
fields, the cosmological evolution of it naturally
leads to the change of the effective fine structure constant.
This can provide an interesting possibility to explain the
observational data of the variation of alpha
($|\Delta \alpha/\alpha|=10^{-5}$-$10^{-6}$)
around the redshift $\tilde{z} \lesssim {\cal O}(1)$.

We adopted a configuration that  branes are located in a
ten-dimensional spacetime with a warp factor $\beta$. Then we
find that the effective fine structure constant is given by
$\alpha=\beta^2 M_s^4/(2\pi V(\phi))$, where $M_s$ is the string
mass scale and $V(\phi)$ is the potential of the field $\phi$.
Since the potential energy at present is related with the Hubble
constant $H_0 \sim 10^{-42}\,{\rm GeV}$, one can estimate the
warp factor $\beta$ by using the present value of alpha
($\alpha=1/137$). This is found to be $\beta \simeq 10^{-61}$
when $M_s$ is the same order as the Planck mass $M_p$. If one
attempts to use the DBI field for the inflation in early
universe, the COBE normalization gives $\beta \simeq 10^{-9}$ for
exponential type potentials \cite{GST}. Therefore we can not use
the single DBI field both for inflation and dark energy unless
some specific mechanism makes the value of $\beta$ smaller after
inflation.

Since $\alpha$ is inversely proportional to $V(\phi)$,
the variation of alpha is dependent on the forms of the DBI potentials.
We have considered three different potentials--(i) exponential:
$V(\phi)=V_0e^{-\mu \phi}$ (Sec.~III),
(ii) inverse power-law: $V(\phi)=M^{4-n}\phi^{-n}$
(Sec.~IV) and (iii) rolling massive scalar:
$V(\phi)=V_0e^{\frac12 M^2\phi^2}$ (Sec.~V).
We performed numerical calculations as well as analytic estimations
in order to confirm the validity of slow-roll approximations.

For the exponential potential transient acceleration occurs when the quantity
$\lambda$ defined in Eq.~(\ref{lam}) is less than unity \cite{CGST},
while the asymptotic solution is dust-like. We found that
both conditions of varying alpha and accelerated expansion
are satisfied for $\mu/H_0=10^{-3}$-$10^{-2}$.
However if we adopt the mass scale $\mu$ motivated by
string theory [see Eq.~(\ref{muexp})], this gives unacceptably
large variation of alpha ($|\Delta \alpha/\alpha| \gg 1$),
which contradicts with observations.

The inverse power-law potential gives rise to an accelerated expansion
at late times for $n \le 2$.
Note that the inverse square potential ($n=2$) corresponds to
the cosmological scaling solution that marks the border of acceleration
and deceleration. We placed constraints on the mass scale $M$
from the information of varying alpha, see Fig.~\ref{mass}.
Although $M$ needs to be much larger than $M_p$ for $n=2$,
this problem is circumvented when $n$ is less than 2.
We have numerically confirmed that it is possible to have
$|\Delta \alpha/\alpha|=10^{-6}$-$10^{-5}$ provided that
the slow-roll parameter $\epsilon$ is nearly constant with
$\epsilon \ll 1$ around the redshift $\tilde{z}={\cal O}(1)$.
Nevertheless this potential is again disfavored observationally
if we use the mass scale $M$ constrained by string theory.

The rolling massive scalar potential leads to the acceleration of the
universe
when the amplitude of the oscillation of $\phi$ is small.
For the mass scale $M$ constrained by string theory ($M=\sqrt{\beta}M_s$)
the field $\phi$ oscillates coherently with a time scale $M^{-1}$
much smaller than $H_0^{-1}$.
If the amplitude $\Phi$ satisfies the condition
$|M^\Phi|=10^{-3}$-$10^{-2}$,
it is possible to obtain $|\Delta \alpha/\alpha|=10^{-6}$-$10^{-5}$ for
the redshift $\tilde{z}<{\cal O}(1)$ together with the condition for
accelerated expansion. Then this potential is favoured observationally
unlike exponential and inverse power-law potentials.

We thus found that the varying alpha provides a powerful tool
to constrain several different DBI potentials motivated by string theory.
It is of interest to extend our analysis to the case in which the constraints
coming from CMB and structure formation are taken into account.


\section*{ACKNOWLEDGMENTS}
We are grateful to T.~Padmanabhan for valuable guidance throughout the work
and D.~V.~Ahluwalia-Khalilova, M.~Alishahiha, H.~Chand, E.~J.~Copeland,
N.~Dadhich and T.~Qureshi for useful discussions.



\end{document}